%Paper: hep-th/9509092
%From: Hugo Compean <Hugo.Compean@fis.cinvestav.mx>
%Date: Sat, 16 Sep 1995 11:22:27 -0600 (CST)
%Date (revised): Tue, 3 Oct 1995 19:56:28 -0600 (CST)

\input phyzzx

\def\IR{{\hbox{{\rm I}\kern-.2em\hbox{\rm R}}}}
\def\IB{{\hbox{{\rm I}\kern-.2em\hbox{\rm B}}}}
\def\IN{{\hbox{{\rm I}\kern-.2em\hbox{\rm N}}}}
\def\IC{{\ \hbox{{\rm I}\kern-.6em\hbox{\bf C}}}}

\def\IZ{{\hbox{{\rm Z}\kern-.4em\hbox{\rm Z}}}}
\def\to{\rightarrow}

\def\underarrow#1{\vbox{\ialign{##\crcr$\hfil\displaystyle
{#1}\hfil$\crcr\noalign{\kern1pt
\nointerlineskip}$\longrightarrow$\crcr}}}
% use of underarrow
%A~~~\underarrow{a}~~~B
%

\def\ltorder{\mathrel{\raise.3ex\hbox{$<$}\mkern-14mu
             \lower0.6ex\hbox{$\sim$}}}
\def\lesssim{\mathrel{\raise.3ex\hbox{$<$}\mkern-14mu
             \lower0.6ex\hbox{$\sim$}}}

%The usage is $X \catquot G$.

\font\Lina=cmbx10 scaled \magstep1
\font\Hugo=cmbx10 scaled \magstep2

\input phyzzx
\overfullrule=0pt
\tolerance=5000
\overfullrule=0pt
\twelvepoint

\twelvepoint

%\rightline{CINVESTAV-FIS 17/95}

%\date{September, 1995}

\titlepage
\title{  \bf FROM PRINCIPAL CHIRAL MODEL TO SELF-DUAL GRAVITY}
\vglue-.25in
\author{Jerzy F. Pleba\'nski, \foot{E-mail: pleban@fis.cinvestav.mx} Maciej
Przanowski\foot{Permanent address: Institute of Physics, Technical University
of L\'od\'z, W\'olcza\'nska 219, 93-005 L\'od\'z, Poland.} and Hugo
Garc\'{\i}a-Compe\'an\foot{E-mail
: compean@fis.cinvestav.mx}}
\medskip
\address{Departamento de F\'{\i}sica
\break  Centro de Investigaci\'on y de
Estudios Avanzados del IPN.
\break Apdo. Postal 14-740, 07000, M\'exico D.F., M\'exico.}
\bigskip
\abstract{  It is demonstrated that the action of SU$(N)$ principal chiral
model leads in the limit $N \to {\infty}$ to the action for Husain's heavenly
equation. The principal chiral model in the Hilbert space $L^2(\Re^1)$ is
considered and it is shown,
that in this case the chiral equation is equivalent to the Moyal deformation of
Husain's heavenly equation. New method of searching for solutions to this
latter equation, via Lie algebra representations in $L^2(\Re^1)$ is given.}
\endpage

\chapter{\bf Introduction}

Recently a great deal of interest has been devoted to the relation between
chiral models and self-dual gravity$^{1-4}$. In particular, Husain$^3$ has
shown that the Ashtekar-Jacobson-Smolin equations describing the self-dual
vacuum metric can be written a
s the principal chiral model equations for the Lie algebra sdiff$(\Sigma^2)$.
This enables him to derive a new form of the self-dual gravity equation which
we call {\it Husain's heavenly equation}.

In the present paper we show that Husain's heavenly equation can be derived
from the action which appears to be the $N \to \infty$ limit of the action for
the SU$(N)$ principal chiral model (Section 2).

Then in Section 3 we consider the principal chiral model in the Hilbert space
$L^2(\Re^1)$.  We prove  that in this case the chiral equation is in fact the
Moyal deformation of Husain's heavenly equation.

Finally, in Section 4 it is shown that given a solution of the principal chiral
equations for some complex Lie algebra ${\cal G}_C$ and a representation of
${\cal G}_C$ into the Lie algebra $\hat{\cal I}$ of linear operators acting in
$L^2(\Re^1)$ one get
s, via the Weyl-Wigner-Moyal formalism the solution of the Moyal deformation of
Husain's heavenly equation. Then, assuming that this solution is analytic with
respect to the deformation parameter $\hbar$ and putting $\hbar \to 0$ one
finds the solution of
 Husain's heavenly equation.

Therefore, in a sense, the looking for the self-dual vacuum metrics consists in
the searching for the Lie algebra representations in the Hilbert space
$L^2(\Re^1)$.

Two natural questions arise

\noindent
(a)-. What is the connection between our method and the twistor method?

\noindent
(b)-. Does the method presented, provide us with a model of the well known, but
still mysterious, procedure$^{5-7}$  su$(N \to {\infty}) \cong {\rm
sdiff}(\Sigma^2)$?

We intend to consider these problems elsewhere.

\chapter{\bf The principal chiral model approach to self-dual gravity}

We deal with {\bf G}-principal chiral model in a simply connected open
submanifold ${\cal V}$ of the two-dimensional Euclidean space $\Re^2$ of
Cartesian coordinates $(x,y)$. Here {\bf G} stands for a matrix real Lie group.
The real Lie algebra of {\bf G}

 will be denoted by ${\cal G}$.

The principal chiral equations can be written as follows
$$\partial_x A_y - \partial_y A_x + [A_x,A_y] = 0, \eqno(2.1a)$$
$$ \partial_x A_x + \partial_y A_y = 0, \eqno(2.1b)$$
where $A_{\mu} \in {\cal G} \otimes C^{\infty}({\cal V}), \mu \in \{x,y\}$, are
the chiral potentials.

{}From $(2.1a)$ one infers that $A_x$ and $A_y$ are of the pure gauge form {\it
i.e.},

$$ A_{\mu} = g^{-1} \partial_{\mu} g, \eqno(2.2)$$
where $g=g(x,y) \in {\rm \bf G}, \  \mu \in \{x,y\}.$

Substituting $(2.2)$ into $(2.1b)$ we get the well known principal chiral
equations

$$ \partial_x\big(g^{-1} \partial_x g \big) + \partial_y \big( g^{-1}
\partial_y g\big) = 0. \eqno(2.3)$$
However, another approach is also available$^8$. Namely, from Eq. $(2.1b)$ it
follows that

$$ A_x = - \partial_y \theta, \ \ \ \ \ {\rm and} \ \ \ \ \  A_y = \partial_x
\theta, \eqno(2.4)$$
where $\theta = \theta(x,y)$ is some ${\cal G}$-valued $C^{\infty}$-real
function on ${\cal V}$, {\it i.e}, $\theta \in {\cal G} \otimes
C^{\infty}({\cal V})$. Inserting then $(2.4)$ into $(2.1a)$ one obtains the
principal chiral equations in the followin
g form

$$ \partial_x^2 \theta + \partial_y^2 \theta + [\partial_x \theta, \partial_y
\theta] = 0. \eqno(2.5)$$
Assume that the Lie algebra ${\cal G}$ of {\bf G} is semisimple. Then by
straightforward calculations one can show that Eqs. (2.5) are the result of the
variational principle$^8$

$$ \delta S_{Ch} = 0,  \ \ \ \ \  S_{Ch} = \int_{\cal V} {\cal L}_{Ch} \  dx
dy$$

$$ {\cal L}_{Ch} := \alpha {\rm Tr} \bigg \{ {1\over 3} \theta [\partial_x
\theta,
\partial_y \theta] - {1 \over 2} \bigg( (\partial_x \theta)^2 + (\partial_y
\theta)^2 \bigg) \bigg \}. \eqno(2.6)$$
where $\alpha > 0$ is some constant. (Compare also with Ref.9).

Consider now the case when
$$ {\rm \bf G} = {\rm SDiff} (\Sigma ^2), \eqno(2.7)$$
where $\Sigma ^2$ is a real two-dimensional flat symplectic manifold of the
symplectic form $\omega = dq \wedge dp$. Then in the case considered we
have$^{5-7}$

$$ {\cal G} = {\rm sdiff}(\Sigma^2) \cong {\rm the \  Poisson \  bracket \
algebra \ on \ } \Sigma ^2. \eqno(2.8)$$
Consequently, $\theta \in {\rm sdiff}(\Sigma ^2) \otimes C^{\infty}({\cal V})$
is a Hamiltonian vector field

$$ \theta = {\partial \Theta \over \partial q} {\partial \over \partial p} -
{\partial \Theta \over \partial q} {\partial \over \partial p}, \eqno(2.9)$$
where $\Theta = \Theta (x,y,p,q)$.
Substituting (2.9) into (2.5) one gets the following equation

$$ \partial^2_x \Theta + \partial^2_y \Theta + \{\partial_x \Theta,
\partial_y \Theta \}_P = {\cal Y}, \eqno(2.10)$$
where ${\cal Y} = {\cal Y}(x,y)$ is some real function and $\{\cdot,
\cdot \}_P$ stands for the Poisson bracket on $\Sigma^2$, {\it i.e.},

$$ \{\partial_x \Theta, \partial_y \Theta \}_P := {\partial (\partial_x
\Theta) \over \partial q} {\partial (\partial_y \Theta) \over  \partial
p} - {\partial (\partial_x \Theta) \over \partial p} {\partial
(\partial_y \Theta) \over  \partial q}. \eqno(2.11)$$

{}From the very definition of $\Theta$ (2.9) one quickly infers that
without any loss of generality the function ${\cal Y}$ can be chosen to
vanish. Finally we arrive at the sdiff$(\Sigma^2)${\it - principal
chiral equation}

$$ \partial^2_x \Theta + \partial^2_y \Theta + \{ \partial_x \Theta,
\partial_y \Theta \}_P = 0. \eqno(2.12)$$
This is exactly {\it Husain's heavenly equation}$^3$ determining the
self-dual metric on ${\cal V} \times \Sigma^2$. (For other heavenly equations
see Refs. 9,10 and 11).
The evident formal similarity between Eqs. (2.5) and Eq. (2.12) may suggest
that this latter equation can be derived from the variational principle
analogous to (2.6). Indeed, by simple computations one finds that Eq. (2.12)
appears to be the Euler-Lagran
ge equation for the following variational problem

$$ \delta S_G = 0, \ \ \ \ \ \ \ \ S_G = \int_{{\cal V} \times \Sigma^2}
{\cal L}_G \ dxdydpdq$$

$$ {\cal L}_G := -{1\over 3} \Theta \{ \partial_x \Theta, \partial_y
\Theta\}_P + {1 \over 2} \bigg( (\partial_x \Theta)^2 + (\partial_y \Theta)^2
\bigg). \eqno(2.13)$$
(In the cases of the first or second heavenly equations$^9$ the Lagrangians
analogous to ${\cal L}_G$ have been considered by Boyer, Finley and one
of us (J.F.P.)$^{10}$. Compare also ${\cal L}_G$ with Ref. 9).

Note that the variational principle (2.13) can be justified as follows:
Let {\bf G} = SU(N) and, consequently, ${\cal G} = {\rm su(N)}$. It is
well known that
$$ {\rm su}(\infty) \cong {\rm sdiff}(\Sigma^2). \eqno(2.14)$$
Then, taking the constant $\alpha$ in (2.6) to be now

$$ \alpha = { (2 \pi)^4 \over N^3}, \eqno(2.15)$$
one can prove that $S_G$ appears to be  an $N \to \infty$ limit of
$S_{Ch}$. This limit can be obtained formally by the substitutions$^9$

$$ {(2\pi)^4 \over N^3} {\rm Tr} (\cdot \cdot \cdot ) \to -
\int_{\Sigma^2} (\cdot \cdot \cdot) dpdq,$$ $$ \theta \to \Theta$$
$$ [\partial_x \theta, \partial_y \theta] \to \{\partial_x \Theta,
\partial_y \Theta \}_P, \eqno(2.16)$$
(About su$(N\to \infty)$ see Refs. 5,6,7 and 9).

\chapter{\bf The principal chiral model in a Hilbert space and the Moyal
deformation of the heavenly equation}

Here we consider the case when the group {\bf G} appears to be the group of
unitary operators acting in the Hilbert space $L^2(\Re^1)$.  Then, ${\cal G}$
is now the Lie algebra of the anti-self-dual operators in $L^2(\Re^1)$. To make
our notation more tra
nsparent we use in the present case the symbols $\hat{\bf U}$ and $\hat{\cal
U}$ for {\bf G} and ${\cal G}$, respectively. The principal chiral equations
(2.5) take now the form of the operator equation

$$ \partial^2_x \hat \theta + \partial^2_y \hat \theta + [\partial_x \hat
\theta, \partial_y \hat \theta] = 0, \eqno(3.1)$$

$$ \hat \theta \in \hat {\cal U} \otimes C^{\infty}({\cal V}).$$

Finally, defining the self-dual operator-valued function $\hat \Theta = \hat
\Theta (x,y)$
$$ \hat \Theta := i {\hbar} \hat \theta, \ \ \ \ \ \hat \Theta  \in {\cal O}
\otimes C^{\infty}({\cal V}), \eqno(3.2)$$
where ${\cal O}$ stands for the set of the self-dual operators in $L^2(\Re^1)$,
we get the principal chiral equation in the Hilbert space $L^2(\Re^1)$ to be

$$ \partial^2_x \hat \Theta + \partial^2_y \hat \Theta + {1 \over i{\hbar}}
[\partial_x \hat \Theta, \partial_y \hat \Theta] = 0. \eqno(3.3)$$
By simple calculations one can quickly show that Eq. (3.3) can be derived from
the following  variational principle (compare with Ref. 9)

$$ \delta S^{(q)}_{Ch} = 0,  \ \ \ \ \ \ \  \ \  S^{(q)}_{Ch} = \int_{\cal V}
{\cal L}^{(q)}_{Ch}  \ dx dy$$

$${\cal L}^{(q)}_{Ch}:= {\rm Tr} \bigg \{ 2\pi {\hbar} \bigg[ - {1 \over
3i{\hbar}} \hat \Theta [ \partial_x \hat \Theta, \partial_y \hat \Theta] +
{1\over 2} \bigg( (\partial_x  \hat \Theta)^2 + (\partial_y \hat \Theta)^2
\bigg) \bigg] \bigg \}$$

$$ = 2\pi {\hbar}  \sum_j <\psi_j| \bigg \{ - {1 \over 3i {\hbar}} \hat \Theta
[\partial_x \hat \Theta, \partial_y \hat \Theta] + {1 \over 2} \bigg(
(\partial_x \hat \Theta)^2 + (\partial_y \hat \Theta)^2 \bigg) \bigg \}
|\psi_j>, \eqno(3.4)$$
where $\{ |\psi_j>\}_{j \in {\bf {\cal N}}}$ is the orthonormal basis in
$L^2(\Re^1)$

$$ <\psi_j| \psi_k> = \delta_{jk}, \ \ \ \ \ \sum_j |\psi_j><\psi_j| = \hat I.
\eqno(3.5)$$

Now we arrive at the point where the Wigner-Weyl-Moyal formalism$^{12-18,9}$
can be applied.
By the {\it Weyl correspondence} one gets the real function on ${\cal V} \times
\Re^2$ defined as follows

$$ \Theta = \Theta(x,y,p,q) := \int_{- \infty}^{+ \infty} < q- {\xi\over 2}|
\hat \Theta | q + {\xi \over 2}> {\rm exp} \big( {ip \xi \over {\hbar}} \big)
d\xi. \eqno(3.6)$$
As $\hat \Theta$ satisfies Eq. (3.3) the function $\Theta$ satisfies the
equation

$$ \partial^2_x \Theta + \partial^2_y \Theta + \{ \partial_x \Theta, \partial_y
\Theta \}_M = 0, \eqno(3.7)$$
where $\{\cdot,\cdot \}_M$ stands for the Moyal bracket {\it i.e.},

$$ \{f_1,f_2\}_M := {1 \over i {\hbar}} (f_1 * f_2 - f_2* f_1) = {2 \over
{\hbar}} f_1 sin({{\hbar}\over 2} \buildrel {\leftrightarrow} \over {\cal P})
f_2, \eqno(3.8)$$
where $\buildrel {\leftrightarrow}\over {\cal P} : {\buildrel
{\leftarrow}\over{\partial} \over \partial q} {\vec  \partial \over \partial p}
- {\buildrel{\leftarrow}\over{\partial} \over \partial p} {\vec \partial \over
\partial q}$, $f_1=f_1(x,y,p,q)$,

$f_2 = f_2(x,y,p,q),$ with the {\it Moyal $*$-product} defined by

$$ f_1*f_2 := f_1 {\rm exp} ({i {\hbar} \over 2}
\buildrel{\leftrightarrow}\over{\cal P})f_2. \eqno(3.9)$$

{}From (3.8) and (3.9) one easily infers that

$$\lim_{{\hbar}\to 0} f_1 * f_2 = f_1f_2 \ \  {\rm and} \ \  \lim_{{\hbar} \to
0} \{f_1,f_2\}_M = \{f_1,f_2\}_P. \eqno(3.10)$$

Therefore, Eq. (3.7) is evidently the {\it Moyal deformation of Husain's
heavenly equation} (2.12).

One is also able to express $S^{(q)}_{Ch}$ and ${\cal L}^{(q)}_{Ch}$ in terms
of $\Theta$ defined by (3.6). A straightforward computation shows that

$$ S^{(q)}_{Ch} = \int_{{\cal V} \times \Re^2} {\cal L}^{(M)}_G \ dx dy dp
dq,$$
$$ {\cal L}^{(q)}_{Ch} = \int_{\Re^2} {\cal L}^{(M)}_G dp dq, $$

$$ {\cal L}^{(M)}_G := - {1 \over 3} \Theta * \{\partial_x \Theta, \partial_y
\Theta \}_M + {1 \over 2} \bigg( (\partial_x \Theta)*(\partial_x \Theta) +
(\partial_y \Theta)*(\partial_y \Theta) \bigg). \eqno(3.11)$$

Thus ${\cal L}^{(M)}_G$ can be considered to be the Lagrangian for Eq. (3.7).
Of course, by (3.10) one has

$$ \lim_{{\hbar} \to 0} {\cal L}^{(M)}_G = {\cal L}_G. \eqno(3.12)$$

It is of some interest to consider in the present case the principal chiral
equations as written in the usual form (2.3).

We now have evidently

$$ \partial_x \big( \hat{g}^{-1} \partial_x \hat{g}\big) + \partial_y
\big(\hat{g}^{-1} \partial_y \hat{g} \big) = 0, \eqno(3.13)$$
where $\hat{g} = \hat{g}(x,y)$ is an $\hat{\bf U}$-valued function on ${\cal
V}$. Eq. (3.13) can be easily found to follow from the variational principle

$$ \delta S'^{(q)}_{Ch} = 0, \ \ \ \ \ \ \ \ S'^{(q)}_{Ch} = \int_{\cal V}
{\cal L}'^{(q)}_{Ch} dx dy$$

$${\cal L}'^{(q)}_{Ch}= - \pi {\hbar}^3 {\rm Tr} \big \{ (\hat{g}^{-1}
\partial_{\mu} \hat{g})(\hat{g}^{-1}\partial_{\mu} \hat{g}) \big \}$$
$$ = \pi {\hbar}^3 {\rm Tr} \big \{ (\partial_{\mu}
\hat{g})(\partial_{\mu}\hat{g}^{-1}) \big \}, \ \ \  \mu \in \{x,y\}
\eqno(3.14)$$
(Summation over $\mu$ !). The Lagrangian (3.14) is well known in the principal
chiral model theory$^{20}$. The constant $\pi {\hbar}^3$ is chosen for further
convenience.

Using the Weyl-Wigner-Moyal formalism one gets

$$S^{\prime (q)}_{Ch} = \int_{{\cal V} \times \Re^2} {\cal L}^{\prime (M)}_G \
dx dy dp dq$$
$${\cal L}^{\prime (q)}_{Ch} = \int_{\Re^2} {\cal L}_G^{\prime (M)} dp dq =
\int_{\Re^2} {\cal L}_G^{\prime \prime (M)} dp dq, $$

$${\cal L}^{\prime (M)}_G := - {{\hbar}^2 \over 2} \big(g^{- \buildrel{*}
\over{1}} * \partial_{\mu} g \big)* \big( g^{- \buildrel {*} \over{1}} *
\partial_{\mu} g \big),$$
$$ {\cal L}^{\prime \prime (M)}_G : = {{\hbar}^2 \over 2} \big(\partial_{\mu} g
\big)*\big(\partial_{\mu} g^{-\buildrel{*} \over{1}} \big), \eqno(3.15)$$
where

$$ g = g(x,y,p,q) := \int_{- \infty}^{+ \infty} <q -{\xi \over 2} | \hat
g| q + {\xi \over 2}> {\rm exp} ({i p \xi \over h}) d\xi \eqno(3.16)$$
and $g^{- \buildrel{*}\over{1}}$ is the Moyal $*$-product inverse of $g$ {\it
i.e.,}

$$ g^{- \buildrel{*} \over{1}} * g = g * g^{- \buildrel{*}\over{1}} = 1.
\eqno(3.17)$$

Moreover, as $\hat g^{\dag} = \hat g ^{-1}$ one has
$$ g^{- \buildrel{*} \over{1}} = \bar g, \eqno(3.18)$$
where the bar stands for the complex conjugation.

In terms  of the function $g = g (x,y,p,q)$ Eq. (3.13) reads

$$ \partial_{\mu} \bigg( g^{- \buildrel{*}\over{1}} * \partial_{\mu} g \bigg) =
0. \eqno(3.20)$$

This last equation is {\it equivalent} to the Moyal deformation of
Husain's heavenly equation (3.7) the correspondence between Eq. (3.7)
and Eq. (3.20) is given by

$$ g^{- \buildrel{*}\over{1}} * \partial_x g = - {1 \over i {\hbar}} \partial_y
\Theta, \ \ \ \ \ \
\ \ g^{-\buildrel{*}\over{1}} * \partial_y g = {1 \over i {\hbar}} \partial_x
\Theta. \eqno(3.21)$$
In terms of $\Theta$ the Lagrangian ${\cal L}_G^{\prime (M)}$ reads

$${\cal L}^{\prime (M)}_G = {1 \over 2}
\big[ (\partial_x \Theta) * (\partial_x \Theta) + (\partial_y
\Theta)*(\partial_y \Theta) \big]. \eqno(3.22)$$
The difference between ${\cal L}^{(M)}_G$ defined by (3.11) and ${\cal
L}'^{(M)}_G$

$${\cal L}^{(M)}_G - {\cal L}^{\prime (M)}_G = - {1\over 3} \Theta *
\{\partial_x \Theta, \partial_y \Theta \}_M, \eqno(3.23)$$
when the relations (3.21) are applied, vanishes. Indeed, one quickly
shows that

$$ \{ \partial_x \Theta, \partial_y \Theta \}_M = \{g^{- \buildrel{*}\over{1}}
* \partial_y
g, g^{- \buildrel{*}\over{1}}*\partial_x g \}_M \equiv 0. \eqno(3.24)$$
Gothering all that, one concludes that {\it Eq. (3.20) with (3.18) give an
equivalent to (3.7) description of the Moyal deformation of Husain's
heavenly equation}.

Finally, we make an important remark. The considerations of Sects. 2 and 3 can
be easily generalized on the case of the {\it complex} Lie algebras ${\cal
G}_C$. In this case the solutions of the principal chiral equations (2.5) are
${\cal G}_C$-valued $C^
{\infty}$-complex functions {\it i.e.} $\theta = \theta(x,y) \in {\cal G}_C
\otimes C^{\infty}({\cal V};{\bf C})$ and  the solutions of Husain's equation
(2.12) or of its Moyal deformation (3.7) are admitted to be complex. Therefore,
consequently, instead
 of $\hat{\cal U}$ one deals with the complex Lie algebra of operators in
$L^2(\Re^1)$. This algebra will be denoted by $\hat{\cal I}$ . Thus we have now
$\hat \Theta \in  \hat{\cal I} \otimes C^{\infty}({\cal V};{\bf C})$ and the
function $\Theta = \Theta(x,y,p,q)$ defined by (3.6) is,
in general, a complex function. Then the
extension of the results on the case of complex coordinates $(x,y,p,q)$ is
automatic.

\chapter{\bf Remarks on looking for the solutions}

The results of the last two sections provide us with a promising method for
searching for the solutions of Husain's heavenly equation (2.12).

Indeed, let
$$ \Phi : {\cal G}_C \to \hat{\cal I} \eqno(4.1)$$
be a Lie algebra homomorphism. This implies that if

$$\theta = \theta(x,y) = \theta_a(x,y) \tau_a
\in {\cal G}_C \otimes C^{\infty}({\cal V}; {\bf C}), \eqno(4.2)$$
where $\tau_a \in {\cal G}_C$, $a=1,..., dim {\cal G}_C$, constitute the basis
of ${\cal G}_C$, is a solution of the principal chiral equations (2.5), then

$$\hat \Theta = \hat \Theta(x,y) = i {\hbar} \theta_a \hat{X}_a \in \hat{\cal
I} \otimes C^{\infty}({\cal V};{\bf C}) \eqno(4.3)$$
$$\hat{X}_a := \Phi(\tau_a)$$
satisfies Eq. (3.3). (Summation over $a$ is assumed!). Consequently, the
function

$$\Theta = \Theta(x,y,p,q) = i{\hbar} \theta_a(x,y) X_a(p,q), \eqno(4.4)$$
where (see (3.6))

$$ X_a = X_a(p,q) := \int_{- \infty}^{+ \infty} < q- {\xi\over 2}| \hat X_a| q
+ {\xi \over 2}> {\rm exp} \big( {ip \xi \over {\hbar}} \big) d\xi,
\eqno(4.5)$$
appears to be a solution of Eq. (3.7). Assume then, that the function $\Theta$
is an analytic function of $\hbar$ {\it i.e.},

$$ \Theta = \sum_{n = 0}^{\infty} {\hbar}^n \Theta_n, \eqno(4.6)$$

$$\Theta_n = \Theta_n(x,y,p,q).$$

As $\Theta$ is a solution of (3.7) and $\lim_{{\hbar} \to 0} \{\cdot,\cdot \}_M
= \{\cdot, \cdot \}_P$ the function $\Theta_0$  fulfills Husain's heavenly
equation (2.12).

It is evident from our previous considerations (see especially the end of \S3)
that if one restricts oneself to the real Lie algebra ${\cal G}$ and, then, to
the homomorphism

$$ \Phi  : {\cal G} \to \hat{\cal U}, \eqno(4.7)$$
then one arrives at the {\it real} solution $\Theta_0$.

Finally, the self-dual metric is defined by$^3$

$$  ds^2= dx \big(\Lambda_{xp} dp + \Lambda_{xq} dq \big) + dy \big(
\Lambda_{yp} dp + \Lambda_{yq} dq \big) + { 1\over \{\Lambda_x, \Lambda_y\}_P}
\bigg[ \big( \Lambda_{xp} dp + \Lambda_{xq} dq \big)^2 + \big(\Lambda_{yp} dp +
\Lambda_{yy} dq\big)^2 \bigg],\eqno(4.8)$$

$$\Lambda \equiv \Theta_0$$
where $ \Lambda_{xp} \equiv \partial_x\partial_p \Lambda$, $\Lambda_{yp} \equiv
\partial_y \partial_p \Lambda$, ...etc.

\section{Examples}

\subsection{The SU(2) Chiral Model}

Here, the solution of the principal chiral equations (2.5) can be written in
the form

$$ \theta = \theta(x,y) = \theta_a(x,y) \tau_a, \ \ \  a=1,2,3$$

$$ \tau_1 = {i \over 2} \pmatrix{ 0 & 1 \cr 1 & 0}, \ \ \  \tau_2 = {i \over 2}
\pmatrix { 0 & -i \cr i & 0}, \ \ \   \tau_3 = {i \over 2} \pmatrix{ 1 & 0 \cr
0 & -1}, \eqno(4.9)$$

$$ [\tau_a,\tau_b] = \epsilon_{abc} \tau_c. $$

We define the Lie algebra homomorphism $\Phi: {\rm su}(2) \to \hat{\cal I}$
by$^{20}$

$$  \Phi(\tau_1) = \hat X_1 := i \beta \hat{q} + {1\over 2{\hbar}} (\hat{q}^2 -
1)\hat{p},$$

$$  \Phi(\tau_2) = \hat X_2 := - \beta \hat{q} + {i\over 2{\hbar}} (\hat{q}^2 +
1)\hat{p}, \eqno(4.10)$$

$$  \Phi(\tau_3) = \hat X_3 := - i \beta \hat{1} - {1\over {\hbar}} \hat{q}
\hat{p},$$
where $\beta \in \Re$  is any constant; $\hat{q}$ and $\hat{p} = -i {\hbar}
{\partial \over \partial q}$ are the position and momentum operators,
respectively. Then, the complex function $\Theta = \Theta (x,y,p,q)$ defined by
(4.4) and (4.5) is a solution
 of Eq.(3.7) {\it i.e.}, of the Moyal deformation of Husain's heavenly
equation.

Inserting (4.10) into (4.5) and, then the result into (4.4) one quickly finds
  $\Theta$ to be of the form

$$ \Theta = {i \over 2} \theta_1 p(q^2 - 1) - {1\over 2} \theta_2 p(q^2 + 1) -i
\theta_3 pq + {\hbar}\cdot (\beta + {1 \over 2}) \cdot (- \theta_1 q - i
\theta_2 q + \theta_3). \eqno(4.11)$$

Consequently, the function $\Theta_0$

$$ \Theta_0 = {i \over 2} \theta_1 p(q^2 -1) - {1\over 2} \theta_2 p (q^2 + 1)
- i \theta_3 pq, \eqno(4.12)$$
appears to be a complex solution of Husain's heavenly equation (2.12). It is
evident, that taking
$$ \beta = - {1\over 2} + { {\gamma}\over {\hbar}}, \ \ \ \ \ {\gamma}\in \Re,
\eqno(4.13) $$
one finds another solution of Eq. (2.12); namely

$$ \Theta'_0 = \Theta_0 + {\gamma} \cdot \big( - \theta_1 q -i \theta_2 q +
\theta_3 \big). \eqno(4.14)$$

\subsection{ The $SL(2;\Re)$ Chiral Model}

Now the solution of the principal chiral equations (2.5) takes the form

$$ \theta = \theta (x,y) = \theta_a(x,y) \tau_a, \ \ \ \ \  a=1,2,3,$$

$$ \tau_1 = {1 \over 2} \pmatrix{ 0 & 1\cr 1 & 0}, \ \ \  \  \tau_2 = {1 \over
2} \pmatrix{ 0 & -1\cr 1 & 0}, \ \ \ \   \tau_3 = {1 \over 2} \pmatrix{ 1 &
0\cr 0 & -1}, \eqno(4.15)$$

$$ [\tau_1, \tau_2] = \tau_3, \ \ \ \ [\tau_2, \tau_3] = \tau_1, \ \ \ \
[\tau_3, \tau_1] = - \tau_2.$$

Following Ref. 21 we define the Lie algebra homomorphism $ \Phi: {\rm
sl}(2;\Re) \to \hat{\cal U}$ by

$$ \Phi(\tau_1) = \hat{X}_1 := {i \over 4} \big( {\hat{p}^2 \over {\hbar}^2} +
{\delta \over \hat{q}^2} - \hat{q}^2\big),$$

$$ \Phi(\tau_2) = \hat{X}_2 := {i \over 4} \big( {\hat{p}^2 \over {\hbar}^2} +
{\delta \over \hat{q}^2} + \hat{q}^2\big), \eqno(4.16)$$

$$ \Phi(\tau_3) = \hat{X}_3 := {i \over 2} \big( {\hat{q} \hat{p} \over
{\hbar}} - {i \over 2}\big),$$
where $\delta \in \Re$ is a constant.

Defining the functions $ X_a = X_a(p,q)$ according to (4.5) and inserting them
into (4.4) we get the real solution of the Moyal deformation of Husain's
equation to be
$$ \Theta = \Theta(x,y,p,q) = {\hbar}^{-1} \big[ -{1 \over 4} (\theta_1 +
\theta_2)p^2 \big] - {1 \over 2} \theta_3 pq +  {\hbar} \big[ - {\delta \over
4}(\theta_1 + \theta_2) q^{-2} + {1 \over 4} (\theta_1 - \theta_2) q^2 \big].
\eqno(4.17)$$

Thus one arrives at the solution which is not an analytic function in $\hbar$
if $\theta_1 + \theta_2 \not= 0$. Therefore, in general, the function

$$ \Theta_0 = - {1 \over 2} \theta_3 pq \eqno(4.18)$$
doesn't satisfy Eq. (2.12).

Assume

$$ \theta_1 + \theta_2 = 0. \eqno(4.19)$$
Then, using the results of Ref. 20 one shows that $\theta_3 = \theta_3(x,y)$ is
of the following form

$$\partial_x \theta_3 = \partial_y F, \ \ \ \ \ \ \ \ \partial_y \theta_3 = -
\partial_x F, \eqno(4.20) $$
where $ F= F(x,y)$ is an arbitrary real function.

Consequently, $\Theta_0$ defined by (4.18) and (4.20) fulfills Husain's
heavenly equation (2.12) but, as now   $\{ \partial_x \Theta_0, \partial_y
\Theta_0 \}_P = 0$,  this solution doesn't define any self-dual metric (see
(4.8)).

\vskip 1truecm
\centerline{\Lina Acknowledgments}
We are indebted to T. Matos for useful discussions on the principal chiral
model. One of us (M.P.) would like to thank the staff of the Department of
Physics at CINVESTAV, M\'exico D.F., M\'exico, for the warm hospitality.
This work is supported by CONACyT and CINVESTAV, M\'exico, D.F., M\'exico.

\vskip 2truecm

\centerline{\Hugo References}

\item{1.} Q-Han Park, {\it Int. J. Mod. Phys.} {\bf A7}, 1415 (1992).
\item{2.} I.A.B. Strachan, {\it Phys. Lett.} {\bf B283}, 63 (1992).
\item{3.} V. Husain, {\it Phys. Rev. Lett.} {\bf 72}, 800 (1994); {\it Class.
Quantum Grav.} {\bf 11}, 927 (1994).

\item{4.} H. Garc\'{\i}a-Compe\'an and T. Matos, Solutions in self-dual gravity
constructed via chiral equations, to appear in {\it Phys. Rev.} {\bf D52}
(1995).
\item{5.} J. Hoppe, {\it Phys. Lett.} {\bf B215}, 706 (1988).
\item{6.} D.B. Fairlie, P. Fletcher and C.K. Zachos, {\it J. Math. Phys.} {\bf
31}, 1088 (1990).

\item{7.} E.G. Floratos, J. Iliopoulos and G. Tiktopoulos, {\it Phys. Lett.}
{\bf B217}, 285 (1989).

\item{8.}C.R. Nappi, {\it Phys. Rev.} {\bf D21}, 418 (1980).

\item{9.} J.F. Pleba\'nski and M. Przanowski, The Lagrangian of self-dual
gravitational field as a limit of the SDYM Lagrangian, submitted to {\it Phys.
Lett.} {\bf A}.
\item{10.} J.F. Pleba\'nski, {\it J. Math. Phys.} {\bf 16}, 2395 (1975).

\item{11.} C.P. Boyer, J.D. Finley III and J.F. Pleba\'nski, Complex
relativity, ${\cal H}$ and ${\cal HH}$ spaces- a survey, in {\it General
Relativity and Gravitation,} Einstein memorial volume, vol.2, ed. A. Held
(Plenum, New York, 1980) pp. 241-281.

\item{12.} H. Weyl, {\it Z. Phys.} {\bf 46}, 1 (1927).
\item{13.} E.P. Wigner, {\it Phys. Rev.} {\bf 40}, 749 (1932).
\item{14.} J.E. Moyal, {\it Proc. Cambridge Phil. Soc.} {\bf 45}, 99 (1949).

\item{15.} J.F. Pleba\'nski, Institute of Physics of Nicolaus Copernicus
University, Toru\'n, preprint No. 69 (1969).

\item{16.} F. Bayen, M. Flato, C. Fronsdal, A. Lichnerowicz and D. Sternheimer,
{\it Ann. Phys. NY} {\bf 111}, 61 (1978); {\it Ann. Phys. NY} {\bf 111}, 111
(1978).

\item{17.} M. Hillery, R.F. O'Connell, M.O. Scully and E.P. Wigner, {\it Phys.
Rep.} {\bf 106}, 121 (1984).

\item{18.} W.I. Tatarskij, {\it Usp. Fiz. Nauk} {\bf 139}, 587 (1983).

\item{19.} N. Sanchez, Exact solutions in gauge theory, general relativity, and
their supersymmetric extensions, in {\it Geometric Aspects of the Einstein
Equations and Integrable Systems,} ed. R. Martini (Springer-Verlag, Berlin,
Heidelberg, 1985) pp. 1-
76.

\item{20.} A. Wawrzy\'nczyk, {\it Group Representations and Special Functions},
 (D. Reidel Publishing Company, PWN, Warszawa, 1984) pp. 310,311.

\item{21} K.B. Wolf, Integral transform representations of SL$(2,\Re)$, in {\it
Group Theoretical Methods in Physics}, Proceedings, Cocoyoc, M\'exico 1980, ed.
K.B. Wolf (Springer-Verlag, Berlin, Heidelberg, 1980) pp. 526-531.

\endpage
\end